%% file: main_2026Mar.tex
\documentclass[aps,prb,preprintnumbers,amsmath,amssymb,superscriptaddress,twocolumn,10pt,iopjournal]{revtex4-2}

\input{macros}

\begin{document}
\raggedbottom
\setlength{\abovedisplayskip}{5pt}
\setlength{\belowdisplayskip}{5pt}
\setlength{\abovedisplayshortskip}{0pt}
\setlength{\belowdisplayshortskip}{0pt}

\title{Microscopic Theory Revealing 
Ising Criticality with Distinct Sublattice Orders in
Pseudospin-$\frac{1}{2}$ Chain}
\author{Mandev Bhullar}
\affiliation{Department of Physics, University of Toronto, Ontario, Canada M5S 1A7}
\author{Philip Richard}
\affiliation{Department of Physics, University of Toronto, Ontario, Canada M5S 1A7}
\author{Hae-Young Kee}
\email[]{hy.kee@utoronto.ca}
\affiliation{Department of Physics, University of Toronto, Ontario, Canada M5S 1A7}
\affiliation{Canadian Institute for Advanced Research, CIFAR Program in Quantum Materials, Toronto, Ontario, Canada, M5G 1M1}
\date{\today}

\begin{abstract}
The one-dimensional transverse Ising model is a paradigmatic example of quantum criticality. In spin-orbit coupled systems, however, effective Ising interactions arise alongside bond-dependent couplings such as Kitaev ($K$) and $\Gamma$ terms in addition to the Heisenberg ($J$) interaction, leading to complex magnetic orders beyond the pure Ising limit.  We first explore how the generic $J-K-\Gamma$ model with four-fold screw symmetry in a spin-orbit-entangled pseudospin-1/2 chain manifests via sublattice order, and then test whether field-driven transitions retain Ising universality. 
We find sublattice-dependent magnetic order below the critical field and a distinct sublattice pattern persisting above it. Despite these complex magnetic order structures, the transition remains in the Ising universality class with central charge $c=1/2$. Our work provides a route to the microscopic Hamiltonian and emergent Ising criticality while allowing microscopic physics of the sublattice orders to manifest at low and high fields. Application to antiferromagnetic Ising materials such as BaCo$_2$V$_2$O$_8$ is also discussed. 
\end{abstract}

\maketitle

\section{Introduction} 
The Ising model, introduced by E. Ising in 1925 \cite{Ising1925} following Lenz’s earlier proposal \cite{Lenz1920}, has been a paradigmatic framework for understanding phase transitions and critical phenomena for nearly a century. Its quantum counterpart, the transverse-field Ising model, is exactly solvable in one dimension via the Jordan–Wigner transformation, providing detailed insights into the ground state, correlations, and a field-driven quantum phase transition \cite{Pfeuty1970}. The critical point belongs to the Ising universality class with central charge $c=1/2$, reflecting its underlying fermionic description. 

Although the theoretical framework has long been established, the experimental search for materials realizing Ising physics only dates back to studies of anisotropic magnets in the 1970s, with rare-earth compounds such as LiHoF$_4$ \cite{Jensen1976} providing some of the first experimental realizations of the transverse-field Ising model \cite{Kogut1979}. Subsequent efforts explored quasi-one-dimensional and layered magnets, although residual interchain couplings often drive these systems toward higher-dimensional ordering \cite{Sachdev2011}.
It was later recognized that the low-energy excitations in candidate materials are not always well described by the simple domain-wall picture of the pure Ising model. This raised the question of whether the underlying microscopic Hamiltonians truly contain a pure Ising interaction, and what additional accompanying interactions are responsible for the dynamics of domain-wall excitations even in the absence of a transverse field.

Recently, several studies have revisited this perspective in the quasi-one-dimensional magnet CoNb$_2$O$_6$, using different frameworks to account for its low-lying excitations even at zero field \cite{Fava2020,fava2020glide,morris2021np,Woodland2023,Gallegos2024}. While earlier works employed symmetry-allowed interactions to model the magnetic dynamics \cite{Fava2020,fava2020glide,morris2021np,Woodland2023,Gallegos2024}, Churchill \emph{et al.} showed that the microscopic Hamiltonian includes Heisenberg ($J$), Kitaev ($K$), and $\Gamma$ interactions, which are not purely Ising in form \cite{churchill2024transforming}. They demonstrated that the spin–orbit entangled wavefunctions generate effective anisotropic couplings described by a $J$–$K$–$\Gamma$ model, analogous to those proposed for two-dimensional (2D) honeycomb magnets \cite{Jackeli2009PRL,Rau2014,Rouso2024RoPP,Matsuda2025RMP}.
Importantly, it was shown that the one-dimensional (1D) geometry leads to qualitatively different physics: the Kitaev interaction in the 2D honeycomb lattice leads to an isotropic interaction, while in 1D, it produces an anisotropic XXZ-type interaction. It also enables domain-wall motion even in the absence of a transverse field, and the $\Gamma$ interaction plays a crucial role in selecting and pinning the moment direction in the ferromagnetic (FM) ordered phase even in the absence of octahedra distortion \cite{churchill2024transforming}.
The generic $J-K-\Gamma$ model, though applied to FM Ising systems like CoNb$_2$O$_6$, is symmetry-based and therefore also applies to antiferromagnetic (AFM) Ising systems such as BaCo$_2$V$_2$O$_8$.

Here, we first ask how non-Ising interactions manifest in the chain with the screw-axis like BaCo$_2$V$_2$O$_8$.
Do they lead to complex magnetic order patterns, and do they modify the field-driven quantum phase transition? While certain perturbations, such as a longitudinal field, are known to gap out the critical point \cite{Zamolodchikov1989,coldea2010quantum}, it remains unclear whether generic $J$–$K$–$\Gamma$ interactions alter the transition or preserve Ising universality.

In this work, we address this question by studying the spin-$1/2$ $J$-$K$-$\Gamma$ model in chains with screw symmetry. 
We show that the system indeed exhibits sublattice-dependent magnetic order induced by the $\Gamma$ interaction, while the Kitaev interaction generates effective XXZ-like anisotropy.  Under an external magnetic field, we show that the field-driven transition remains governed by Ising criticality with central charge $c=1/2$, despite non-Ising microscopic interactions generate two distinct order patterns below and above the transition. The high-field phase with distinct sublattice order is adiabatically connected to the polarized phase, implying the absence of topological solitons above the transition.  The microscopic theory also provides the origin of the phenomenological site-dependent g-factor widely adapted for BaCo$_2$V$_2$O$_8$\cite{Kimura2007}. 

In what follows, we first derive the microscopic Hamiltonian for a chain with a four-fold screw axis running along the chain direction in Sec. \ref{microH}. We then analyze the classical magnetic ordering pattern in the presence of an external field in Sec. \ref{classanalysis}. The quantum phase transition under a transverse field obtained by the density matrix renormalization group (DMRG) method \cite{White1992} is presented in Sec. \ref{DMRGtransition}, and the analysis of the entanglement entropy leading to the conclusion of Ising criticality in presented in Sec. \ref{entangleentro}. In Sec. \ref{BCVOapp}, we discuss the application of our theoretical results to BaCo$_2$V$_2$O$_8$, and conclude with a summary and discussion in Sec. \ref{sumandout}.

\section{Microscopic Hamiltonian for a chain with four-fold screw symmetry \label{microH}}
\begin{figure}
    \includegraphics[width=0.76\linewidth]{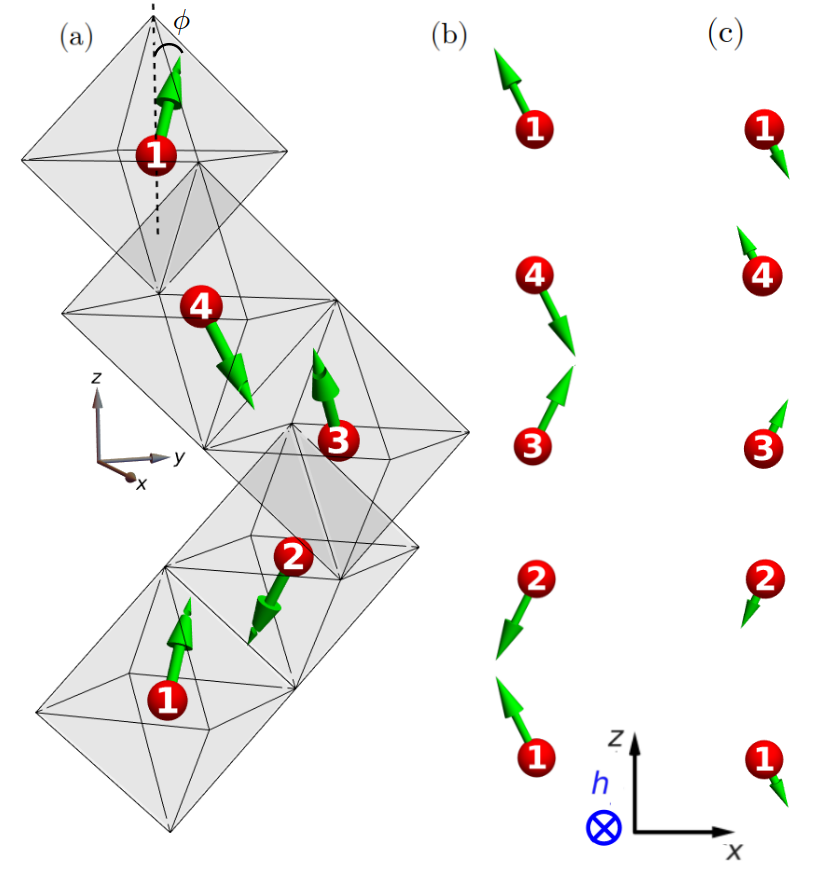}
    \caption{(a) Structure of the spin-1/2 (red circles) screw chain with the zero-field moment directions. The $xyz$ coordinate system is aligned with the crystallographic $abc$ axes. 
    $\alpha=1,2,3,4$, which denote the four sublattices, are shown in each octahedron. The green arrows represent the magnetic moments in the $zx$-plane under a magnetic field \(h\) applied along the $y$-axis for (b) $h < h_{c}$ and (c) $h > h_{c}$ where $h_{c}$ is the critical field. Here, the length of the arrow and the tilt angle $\phi$ are exaggerated for visibility. \label{fig:structure}  }
\end{figure}

We consider the pseudospin \(J_{\mathrm{eff}} = 1/2\) Kramer’s
doublet forming a four-fold screw chain rotating along the crystallographic \(c\)-axis, as shown in Fig. \ref{fig:structure}(a). They are linked by edge-sharing oxygen octahedra which define four local cartesian coordinate systems in a unit cell denoted by $(x_\alpha,y_\alpha,z_\alpha)$ with $\alpha = 1,2,3,4$ representing repeating four-sites. 
In what follows, the effective $J_{\rm eff}=1/2$ is represented by $S=1/2$ for simplicity. 

Using this local coordinate system with ideal octahedra, the n.n. generic 
$S=1/2$ exchange model for each n.n. x- and y-bond running along the screw chain is given by the \(J K \Gamma\) model \cite{Rau2014,Liu2018PRB,Sano2018PRB,churchill2024transforming,Winter_2022,Rouso2024RoPP,Matsuda2025RMP} 
\begin{equation}
 H_{\langle ij \rangle}^{\gamma} =
     J \mathbf{S}_i \cdot \mathbf{S}_j + K S_i^{\gamma} S_j^{\gamma} + (-1)^{\lfloor \frac{i-1}{2}\rfloor} \Gamma \big( S_i^{\beta} S_j^{z} + S_i^{z} S_j^{\beta} \big),
     \label{LocalHamiltonian}
\end{equation}
where $\gamma = x/y$ and $\beta = y/ x$, respectively. The sign structure of the $\Gamma$ term stems from the $\pi/2$ rotation about the screw axis; see the full derivation in the Supplementary Material (SM).
 
Including the four-sublattice dependence, the full Hamiltonian is given by
\begin{eqnarray}
H &=& \sum_{j = 1}^{N} \Bigg\{ J \Bigg[  \epsilon  \big(   S_{j}^x S_{j + 1}^x  + S_{j}^y S_{j + 1}^y   \big) +   S_{j}^z  S_{j + 1}^z   \Bigg] \nonumber\\
&&+\frac{K}{2} \Bigg[ (-1)^{j} \big(    S_{j}^x  S_{j + 1}^x  - S_{j}^y  S_{j + 1}^y  \big) \Bigg]\nonumber\\
&& - \Gamma  \Bigg[ \cos \bigg( \frac{\pi j}{2} \bigg)   \big( S_{j}^{x} S_{j+1}^{z} +  S_{j}^{z}  S_{j + 1}^{x}\big)  \nonumber\\
&& -  \sin \bigg( \frac{\pi j}{2} \bigg)    \big( S_{j}^{y} S_{j + 1}^{z} + S_{j}^{z} S_{j + 1}^{y}\big)  \Bigg] \Bigg\},
\label{ZeroFieldGlobalHamiltonian}
\end{eqnarray}
where $\epsilon \equiv  1 + \frac{K}{2 J}$ 
and \(N\) denotes the number of sites. 
The first line of Eq. \eqref{ZeroFieldGlobalHamiltonian} is the XXZ model, where the anisotropy parameter $\epsilon$ is determined by the ratio of the Heisenberg $J$ and Kitaev $K$ interactions\cite{churchill2024transforming}.

We are interested in the AFM Ising dominant regime, $J>0$ and $\epsilon < 1$, 
(i.e., $K<0$) leading to the dominant AFM Ising interaction in the XXZ terms from our Hamiltonian given by Eq. \eqref{ZeroFieldGlobalHamiltonian}. Thus, the system is described by a dominant FM Kitaev interaction and sub-dominant AFM Heisenberg interaction. We will see in Sec. \ref{classanalysis} that a finite $\Gamma$ interaction not only gives rise to moment tilting from the $c$-axis at zero field but also generates distinct sublattice magnetic orders in low- and high-field regimes.

\section{Classical model and transition analysis \label{classanalysis}}
We first consider the zero-field ordered phase before turning to the field-induced transition.   
We expect the moments \(\mathbf{M}_j \equiv \langle \mathbf{S}_j \rangle\) 
with larger \(z\)-components to display Néel order, i.e. \(M_j^z = (-1)^{j+1} m_z^0\). 
The \(\Gamma\) interaction then behaves as an effective staggered field \( \sim \Gamma  \sum_{j = 1}^{N} (-1)^j \big[ \cos \big( \pi j/2 \big)   \big( S_{j}^{x}  - S_{j + 1}^{x}\big)  - \sin \big( \pi j/2 \big)   \big( S_{j}^{y}  - S_{j + 1}^{y}\big)  \big] \),
leading to a site-dependent $g$-tensor even in the absence of the external field.
In the presence of the alternating component with a FM Kitaev interaction \(K < 0\) and the $\Gamma$ interactions,
the moments have the following sign configurations
\begin{equation}
\mathbf{M}_j [\mathbf{h} = \mathbf{0}] = \begin{Bmatrix}
      \sqrt{2} \cos \big( \frac{2j + 1}{4} \pi
\big) m_x^0, \\  \sqrt{2} \cos \big( \frac{2j - 1}{4} \pi
\big) m_y^0, \\  (-1)^{j+1} m_z^0 
\end{Bmatrix}, \label{ZeroFieldMomentDirection}    
\end{equation}
where $m^0_{x}$ and $m^0_z$ are given by
\begin{eqnarray}
m^0_{x} &=& m^0_y =\frac{\sqrt{8\Gamma^2 + (2J + K )^2} - (2J + K )}{4 \Gamma} m_z^0, \nonumber\\
m^0_z &=& \frac{1}{2\sqrt{2}} \sqrt{1 + \frac{2J + K}{\sqrt{8 \Gamma^2 + (2J + K )^2}}}.   \label{SMmxymzDefinitions}  
\end{eqnarray}
The tilting angle is then given by 
 $\phi = \arctan\left(\frac{\sqrt{2} m_{x}}{m_{z}} \right).$
Eq. \eqref{SMmxymzDefinitions} shows that the $\Gamma$ interaction is responsible for the four-sublattice pattern and moment-tilting at zero field, since both $m^0_x$ and $\phi \rightarrow 0$ as $\Gamma \rightarrow 0$.
The detailed classical analysis is provided in the SM. 

Note that the sign pattern of ${\bf M}_j$  exhibits a four-sublattice periodicity. It is useful to group the sites $j$ into a repeating four-site pattern: $4n+1$, $4n+2$, $4n+3$, $4n+4$, where $n$ indexes the unit cell. Using this notation, the four-site pattern of $M^z_n$ can be written as $M_n^z=(+m_z^0,-m_z^0,+m_z^0,-m_z^0)$ whereas $M^x_n = (-m_x^0,-m_x^0,+m_x^0,+m_x^0)$, as shown by the green arrows in Fig. \ref{fig:structure}(b) in the $z-x$ plane.
$M_j^y$ is then determined by the four-fold screw rotation around the $z$-axis, i.e., $M_n^y = (+m_y^0, -m_y^0, - m_y^0,  +m_y^0)$.
The $\Gamma$ interaction thus generates magnetizations in the $x$--$y$ plane 
that rotate by $90^\circ$ as one moves along the $z$-axis, consistent 
with earlier studies employing a site-dependent $g$-tensor~\cite{Kimura2007}, as we will show below.
However, because 
the site-dependent $g$-tensor requires a finite external magnetic field, 
it does not accurately describe the ordered state at $h = 0$. 

Under a field $h > 0$ applied along the $y$-axis, assuming no broken translational symmetry (i.e. repeating periodic 4-site unit cells), we solve for the classical ground state numerically under the constraint \((S_j^{x})^2 + (S_j^{y})^2 + (S_j^{z})^2 = S^2\) with $S=1/2$ and initial ansatz given by Eqs. \eqref{ZeroFieldMomentDirection} and \eqref{SMmxymzDefinitions} (see the SM for details). Under this theory, we observe a transition at $h = h_c$. 
At lower fields $h < h_c$, we observe the following magnetization pattern:
\begin{eqnarray}
   M_n^x &=& (-m^\prime_x, -m_x^{\prime\prime},+m_x^{\prime\prime},+m_x^\prime)\nonumber\\
    M_n^y &=&(+m_y^\prime,+m_y^{\prime\prime},+m_y^{\prime\prime},+m_y^\prime)\nonumber\\
   M_n^z &=& (+m_z^\prime,-m_z^{\prime\prime},+m_z^{\prime\prime},-m_z^\prime)\ ,
    \label{lowfieldgs1}
\end{eqnarray}
where $m_z^\prime,m_z^{\prime\prime},m_x^\prime,m_x^{\prime\prime}>0$, and $m_y^\prime,m_y^{\prime\prime}$ respectively have negative and positive sign at very low field but continuously evolve until both become positive after a certain field strength. After this field strength, the coefficients obey the inequalities $m_y'' < m_y'$, $m_z' < m_z''$ and $m_x' < m_x''$.

At higher fields $h > h_c$, $M_j^y$ becomes uniform, while $M_j^z$ and $M_j^x$ obey new 
pattern:
\begin{eqnarray}
    M_n^x &=& (+m_x, -m_x,+m_x,-m_x)\nonumber\\
    M_n^y &=&(+m_y,+m_y,+m_y,+m_y)\nonumber\\
M_n^z &=& (-m_z,-m_z,+m_z,+m_z)\ ,\,
\label{highfieldgs}
\end{eqnarray}
where $m_y, \, m_z, \, m_x > 0$, but $m_y > m_z > m_x$ for all field values above the transition. The full classical evolution of the sign patterns of $\mathbf{M}_n$ under the field $h$ is shown in the SM. As $h \rightarrow \infty$, $m_y \rightarrow 0.5$ and $m_z, m_x \rightarrow 0$. Hence, the high-field phase is adiabatically connected 
to the fully polarized state, despite the nontrivial sublattice 
magnetizations. The sublattice-dependent magnetic patterns are illustrated in Fig. \ref{fig:structure} (b) and (c) for the low- and high-field regions.
The summary of the sublattice magnetization patterns below and above $h_{c}$ is also given in Table \ref{tab:signpatterns}.

\begin{table}[t]
\centering
\begin{tabular}{c|c|c|c}
\hline\hline
Field regime & $M_n^x$ & $M_n^y$ & $M_n^z$ \\
\hline\hline
$h=0$ & $(-\;,-\;,+\;,+)$ &
$(+\;,-\;,-\;,+)$ & $(+,-,+,-)$ \\
\hline
$h < h_{c}$ &
$(-\;,-\;,+\;,+)$ &
$(+\;,+\;,+\;,+)$ & $(+,-,+,-)$\\
\hline
$h > h_{c}$ &
$(+\;,-\;,+\;,-)$ &
$(+\;,+\;,+\;,+)$ & $(-,-,+,+)$ \\
\hline
\end{tabular}
\caption{Sign structure of the four sublattice magnetizations 
$M_n^x$, $M_n^y$, and $M_n^z$ for the low- and high-field ordered states. Note that the $y$-component sublattice structure present at $h = 0$ smoothly evolves into a uniform sign as the field $h$ increases, but the magnitude of the sublattice pattern is shown in Fig. \ref{fig:onsitemagnetizations}.
As $h \rightarrow \infty$, the sublattice structure of ${\bf M}_n$ remains unchanged, while the magnitude of the moments $M_n^x$ and $M_n^z$ approach zero.}
\label{tab:signpatterns}
\end{table}

\section{Quantum phase transition under a magnetic field \label{DMRGtransition}} 
We now consider the application of a magnetic field $h$ applied along the $y (\parallel {\hat b})$-axis in the quantum model given by Eq. \ref{ZeroFieldGlobalHamiltonian}. By symmetry, a field applied along the  $x$-axis gives the same result. 
To examine the phase transition and field-induced phases under a field, we add the Zeeman term  $H_{\rm ext} = - g\mu_B \sum_i {\bf h} \cdot {\bf S}_i$, where $g$ is the g-factor and ${\bf h} = h \; {\hat y}$.
Given the ideal octahedra environment, we incorporate \(h\) using 
an isotropic Landé \(g\)-factor of \(g = 4\) \cite{Liu2020PRL} and set $g \mu_B \equiv 1$. 

\begin{figure}[t]
    \centering
   \includegraphics[width=1.01\linewidth]
  {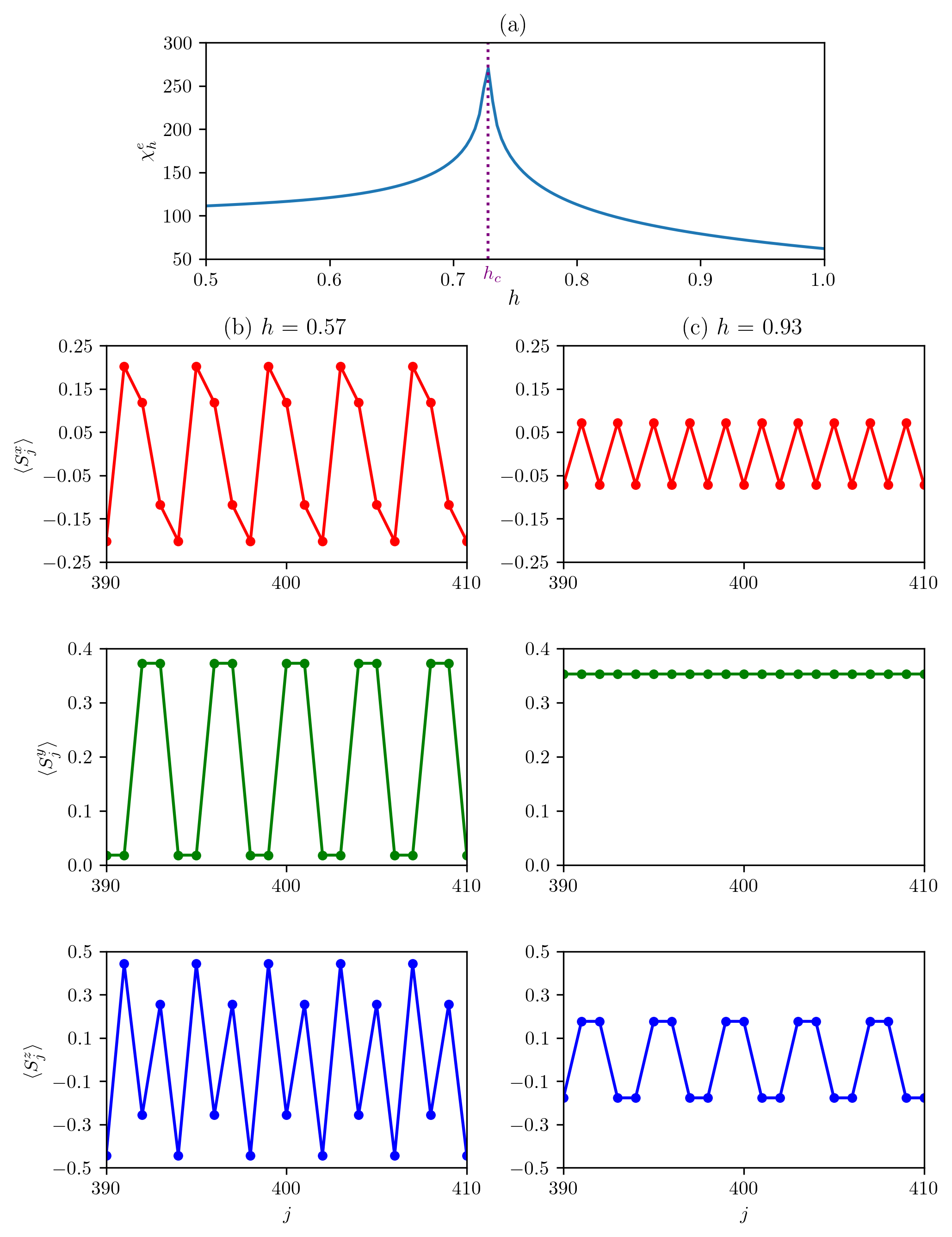}
    \caption{DMRG results for $J=1$, $K = -1.18$ and $\Gamma = 0.7$, obtained for a $N=800$-site chain. (a) The susceptibility $\chi_e^h$ indicating the transition at $h_c$ (marked by the purple dashed line). The on-site magnetizations $\langle S^{\gamma}_j \rangle$ with $\gamma = x,y,z$ in the middle region of the chain showing distinct four-sublattice patterns in two representative regions: (b) low-field ($h=0.57$), and (c) high-field ($h=0.93$) regions. 
    }
    \label{fig:onsitemagnetizations}
\end{figure}

To obtain the critical field where a transition occurs, we employ the DMRG method and study the magnetic susceptibility \(\chi^{e}_{h} = -\partial^2 e_0/\partial h^2\) where \(e_0\) is the energy per site \cite{bhullar2025field}.
We set $J=1$, $\epsilon = 0.41$, and $\Gamma =0.7$. The results are not sensitive to the choice of the parameters, as long as $J>0$, $K<0$, and $|K| < 2J$ (dominant AFM Ising limit) with a finite $\Gamma$.
The result is shown in Fig. \ref{fig:onsitemagnetizations}(a), which shows a sharp feature at $h = h_{c}$ denoted by the purple dashed line. 
Figs. \ref{fig:onsitemagnetizations}(b) and (c) show the on-site magnetizations 
$\langle S^{x/y/z}_{j} \rangle$ in the low-field regime $h < h_{c}$ and high-field regime $h > h_{c}$, respectively.

We find that the magnetizations in the low- and high-field regimes have the same sublattice-dependent pattern presented in the classical model, whose sign is summarized in Table \ref{tab:signpatterns}. We proceed to discuss the ground state degeneracy in each regime.

Since the field is applied along the \(y\)-axis, both time-reversal and screw-axis symmetries are lost. However, the composition of a 2-fold screw-axis and time-reversal, $2_1\times T$, is preserved. This action has the effect of sending $j\to j+2$ and $S^z\to -S^z$, which one can verify leaves the Hamiltonian unchanged. The ground state in the low-field phase $h < h_{c}$ maps to another ground state under this transformation. Indeed, at small field, the ground state has a magnetization configuration given by Eq. \eqref{lowfieldgs1}, obtained by the classical model. Applying the $2_1 \times T$ operator to the ground state gives a new magnetization pattern with the same energy:
\begin{eqnarray}
    M_n^x &=&(+m_x^{\prime\prime}, +m_x^\prime,-m_x^\prime,-m_x^{\prime\prime})\nonumber\\
  M_n^y &=&(+m_y^{\prime\prime},+m_y^\prime,+m_y^\prime,+m_y^{\prime\prime})\nonumber\\
  M_n^z &= & (-m_z^{\prime\prime},+m_z^\prime,-m_z^\prime,+m_z^{\prime\prime})\ .
\end{eqnarray}
Hence, we have two different yet degenerate states, leading to topological solitons and domain excitations between the two different ground states. 

On the other hand, in the high-field phase $h > h_c$, the magnetization pattern is instead given by Eq. \eqref{highfieldgs}, consistent with the result obtained by the classical theory.
This high-field phase exhibits a distinct alternating
sign pattern of the sublattice magnetization as sketched in Fig. \ref{fig:structure}(c). 
Notably, this alternating sign pattern is reversed between $M_x$ and $M_z$, relative to
the ordered state below $h_{c}$. 
Furthermore, since the magnitudes
$m_x$ and $m_z$ are uniform across the sublattices, if we now apply the $2_1\times T$ on this state, we end up with the same state, hence we have a unique ground state for $h> h_c$, which is adiabatically connected to the polarized phase.

\section{Ising criticiality: Entanglement entropy and central charge \label{entangleentro}} 
Let us now investigate the nature of the transition at $h = h_c$.
To understand the nature of the critical point $h_{c}$, we investigate the spin-spin correlator and entanglement entropy. We compute the longitudinal spin-spin correlator
\begin{equation}
\langle S_1^l S_j^l \rangle, \, j = 4n + 1, \,
n = \{ 0, ..., N_c \}, 
\label{longitudinalcorrelator}
\end{equation}
where $S_{j}^l = \frac{1}{\sqrt{2}} \sin(\phi) \big(- S_{j}^x + S_{j}^y \big) + \cos(\phi) S_{j}^z$ for the sublattice $\alpha =1$, $\phi$ is the angle from the $c$-axis as shown in Fig. \ref{fig:structure}(a), and $N_c$ is the number of unit cells.

\begin{figure}[t]
    \centering
   \includegraphics[width=0.9\linewidth]{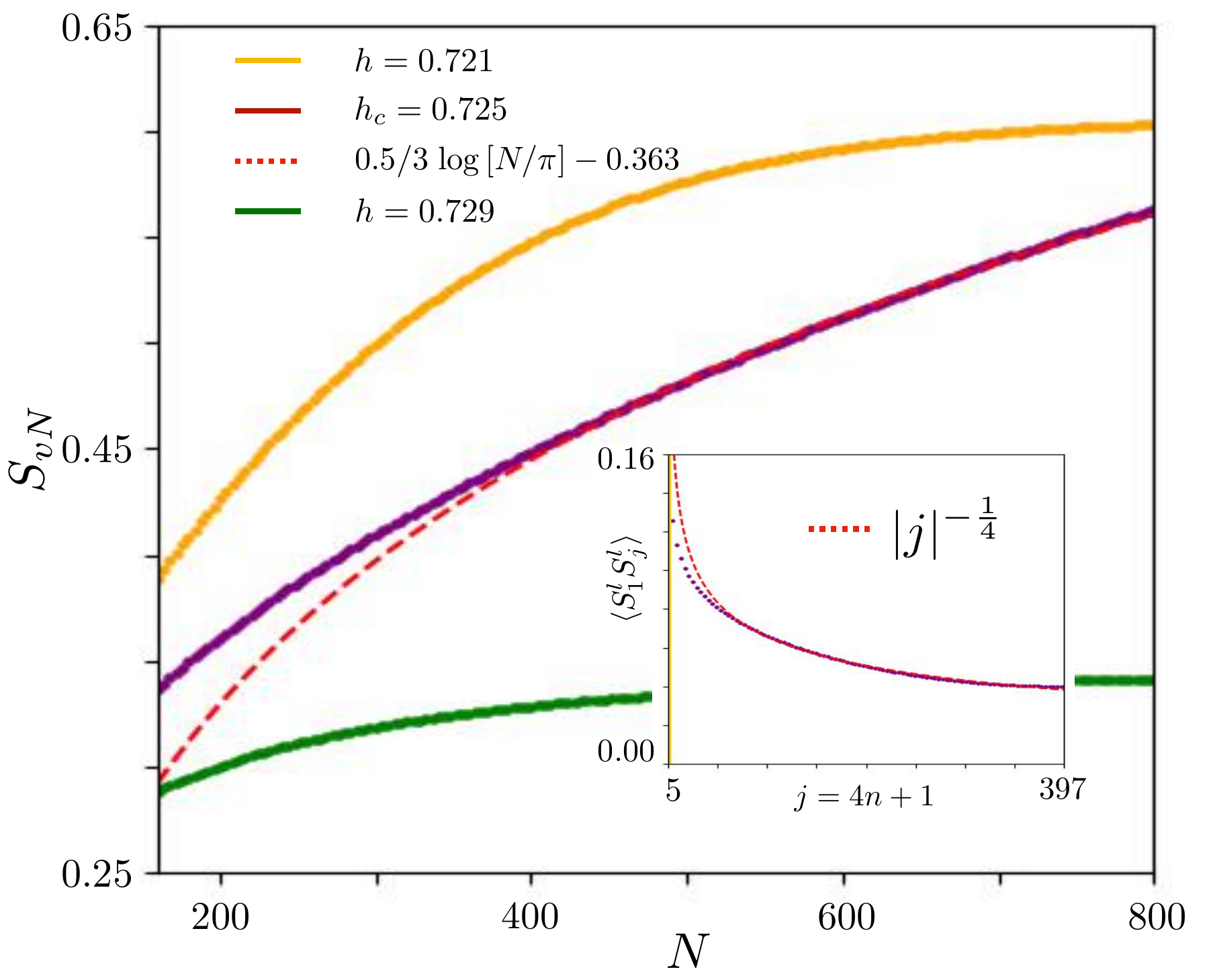}
    \caption{DMRG results
    for the von Neumann entanglement entropy $S_{vN}$ as a function of the chain size $N$ for various values of $h$ near $h_{c} \, (= 0.725)$. At $h = h_{c}$, it scales as $\ln(N/\pi)$ with central charge $c = 1/2$, indicating the Ising critical point, 
whereas its deviation at other fields indicates a finite gap. The inset shows the longitudinal spin--spin correlator $\langle S_1^l S_j^l \rangle$ for the $\alpha=1$ sublattice at $h = h_{c}$, showing power-law behavior with the exponent $\eta=1/4$.
}
    \label{fig:CC_correlator}
\end{figure}

We also determine the von Neumann entropy 
\begin{equation}
S_{vN} = - \mathrm{Tr}\big( \rho_{N'} \ln \rho_{N'} \big),
\label{vonNeumannEntropy}    
\end{equation}
where \(\rho_{N'}\) is the reduced density matrix of a subsystem of length \(N'\) \cite{Sorensen2021PRX, mitra2025quantum} for a chain of length \(N\). We choose \(N' = N/2 - 1\), i.e, the subsystem size is half the chain length. At a critical point, where the spectral gap closes, \(S_{vN}\) scales as
\begin{equation}
S_{vN} \simeq \frac{c}{3} \ln \bigg[ \frac{N}{\pi} \sin \bigg( \frac{\pi N'}{N} \bigg)\bigg] + c' \simeq  \frac{c}{3} \ln \bigg[ \frac{N}{\pi}\bigg] + c',
\label{vonNeumannEntropyCriticalPoint}   
\end{equation}
where \(c\) is the central charge of the underlying conformal field theory, and \(c'\) is a nonuniversal constant \cite{mitra2025quantum}. We show  \(S_{vN}\) for $h$ near $h_{c}$ in Fig. \ref{fig:CC_correlator}. From the purple curve at $h_{c}$, we find the central charge \(c = 0.5\).
The inset shows \( \langle S_1^l S_j^l \rangle \) given by Eq. \eqref{longitudinalcorrelator} at $h_{c}$. Notably, $h_{c}$ is a critical point where the the longitudinal spin-spin correlator 
decays as a power law:
\( \langle S_1^l S_j^l \rangle  \sim |j|^{-\eta}\) where \(\eta = 0.25(1)\). This is consistent with the 2D Ising critical point \cite{Montroll1963_Ising_Correlations,Wu1976_Ising_SpinSpin}.

\section{Application to B\MakeLowercase{a}C\MakeLowercase{o}$_2$V$_2$O$_8$ \label{BCVOapp}}

Among the extensively studied 1D Ising candidate materials, the quasi-one-dimensional compound  
BaCo$_2$V$_2$O$_8$ has emerged as a particularly important example
\cite{Wichmann1986_BaCo2V2O8,He2005crystal,He2005Field,He2006Anisotropy,
Kawasaki2011PRB,kimura2013collapse,Niesen2013PRB,Ideta2013NMR,
faure2018topological,halati2023repulsively}.
In these materials, Co$^{2+}$ ions form chains running along a certain direction.
For example, in BaCo$_2$V$_2$O$_6$, the chains runs along the fourfold
$c$-axis of a tetragonal lattice
\cite{He2006Anisotropy,Kawasaki2011PRB,Mansson2012PhysProcedia,
Kawasaki2014JPSCP,Wang2018PRL_QC,Faure2019PRL_ConfinedSpinons},
as illustrated in Fig.~\ref{fig:structure}(a).

To derive a microscopic theory for 
screw-chains made of magnetic ions such as BaCo\(_2\)V\(_2\)O\(_8\), we first recall the atomic wavefunction of Co\(^{2+}\), which gives rise to 
pseudospin-1/2 through the interplay of Hund’s coupling and spin-orbit coupling. A Co\(^{2+}\) ion with a 3\(d^7\) electron configuration surrounded by an octahedral cage of oxygen atoms generates a cubic crystal field that splits the \(d\)-orbital manifold into \(t_{2g}\) and \(e_{g}\) states, separated by an energy \(\Delta_c\). Due to a large Hund’s coupling \(J_H\) (\(J_H > \Delta_c\)), the Co\(^{2+}\) ion forms a
high-spin \(t^5_{2g}e^2_{g}\)
electron configuration and a 12-fold degenerate \(L_{\rm total} = 1, \, S_{\rm total} = 3/2\) subspace is further split by spin-orbit coupling resulting in a low-energy, pseudospin \(J_{\mathrm{eff}} = 1/2\) Kramer’s
doublet \cite{churchill2024transforming, Liu2018PRB, Sano2018PRB, Liu2020PRL}.

At zero magnetic field, BaCo$_2$V$_2$O$_8$ exhibits AFM order with
strong Ising-like anisotropy \cite{He2005crystal}, where the ordered moments are
slightly tilted away from the $c$ axis by a small angle $\phi$
\cite{kimura2013collapse,Wang2018PRL_QC,faure2018topological,
Faure2021PRR,cui2019quantum,halati2023repulsively}.
Under a magnetic field applied along the $[100]$ direction, thermal expansion
and magnetostriction measurements revealed a transition from the canted
AFM phase to a disordered state
\cite{kimura2013collapse,Niesen2013PRB,Okutani2021JPSJ,
Faure2021PRR,Wang2018PRL_QC},
suggesting an effective description in terms of weakly coupled spin-$1/2$
Ising chains.

Despite this apparent simplicity, unusual field-dependent spin excitations have
motivated extensive studies beyond the pure Ising model.
A phenomenological, site-dependent anisotropic $g$-tensor was introduced to
account for effective staggered fields perpendicular to both the Ising axis and
the applied transverse field
\cite{kimura2013collapse,faure2018topological,zou20218,halati2023repulsively}.
This description captures several experimental features and led to the proposal of dual solitonic excitations below and above the critical field, described by a double sine-Gordon theory with a topological transition
\cite{faure2018topological,giamarchi1988theory,Lecheminant:2002va}.

As shown in Sec. \ref{microH}, the microscopic model for the ideal octahedra cage is then given by the Hamiltonian, Eq. \eqref{ZeroFieldGlobalHamiltonian}.
Based on the AFM  order observed in BaCo$_2$V$_2$O$_8$, the dominant interaction is a FM Kitaev term,
supplemented by AFM Heisenberg and small but finite $\Gamma$ couplings. 
These couplings naturally generate a staggered internal field even in the absence of an external magnetic field that lead to sublattice patterns.
The two-fold ground state degeneracy is lifted at the critical field $h_c$, the transition is not topological in nature, but instead falls into the Ising universality class, demonstrating that the long-distance physics is insensitive to the detailed bond-dependent
structure responsible for the complex magnetic patterns both below and above the critical field.

The high-field magnetization structure has the following structure:
\begin{eqnarray}
M_j^x &= & m_x (h) (-1)^{j + 1}, \, M_j^y =  m_y (h) , \nonumber\\
M_j^z &=&  m_z (h) \cos \bigg( \frac{2j + 1}{4} \pi \bigg),
\label{SMJKGammaHighFieldMagnetization}
\end{eqnarray}
where the field dependent functions $m_x (h)$, $m_y (h)$, and $m_z(h)$ are positive and satisfy the limits \(m_x (h \rightarrow \infty) \rightarrow 0 \), \(m_z (h \rightarrow \infty) \rightarrow 0 \) and \(m_y (h \rightarrow \infty) \rightarrow 0.5\). The above form leads to a site-dependent $g$-tensor. Thus, the $\Gamma$ interaction provides a microscopic origin of the effective $g$-tensor corresponding to the site-dependent phenomenological $g$-tensor used previously. However, we found that the phase above $h_c$ exhibits a unique ground state. 
Accordingly, the $x$- and $z$-component sublattice magnetizations eventually vanish as $h$ becomes large, in contrast to the phenomenological $g$-tensor description. This state is adiabatically connected to the polarized state, and thus our microscopic theory does not support a topological soliton above $h_c$, in contrast to earlier proposals \cite{faure2018topological,Takayoshi2018PRB}. 

Note that the $g$-tensor magnitudes cannot be compared directly, as octahedral distortions may further enhance the anisotropy which is relevant for the high-field regime, and a quantitative assessment of this effect is left for future work. The octahedral distortion also generates additional exchange interactions such as $K^\prime$, $\Gamma^\prime$, and $\Gamma^{\prime\prime}$ \cite{churchill2024transforming}. However, the Ising criticality with central charge $c=1/2$ is expected to persist, as the $Z_2$ symmetry discussed above remains preserved despite the distortion.

Analysis of the transition in the FM Ising-like material CoNb$_2$O$_6$ also reveals Ising criticality. In particular, adopting the microscopic model that was derived in Ref. \cite{churchill2024transforming}, we show in the SM that at the critical field $h_c$, the susceptibility $\chi^e_h$ reveals a second-order transition, the von Neumann entropy $S_{vN}$ shows a central charge $c = 0.5$ and the longitudinal spin-spin correlator $\langle S_1^l S_j^l \rangle$ decays as a power law:
$ \sim |j|^{-0.25}$, all of which are consistent with the findings of this work.

\section{Summary and Outlook \label{sumandout}}
In summary, 
within the generic $J-K-\Gamma$ Hamiltonian, we show that the Kitaev interaction leads to the Ising anisotropy and the off-diagonal $\Gamma$ interaction 
naturally alternates its sign across the four sublattices as a direct consequence of 
the $\pi/2$ screw-chain crystal structure, responsible for the sublattice-dependent magenetic orders below and above the transition field.  

Despite this rich sublattice-dependent magnetic structure, a key question is whether 
such microscopic details influence the long-wavelength critical behavior under an 
external field. Our microscopic theory reveals the robust critical point, consistent with $Z_2$ symmetry (i.e., $2_1 \times T$) discussed above. DMRG calculations establish that the transition at $h_{c}$ remains 
governed by Ising criticality, as evidenced by power-law spin--spin correlations and a 
central charge $c=1/2$ extracted from the von Neumann entanglement entropy.

Applying our theory to an Ising candidate, BaCo$_2$V$_2$O$_8$, we also provide the microscopic origin of the phenomenological 
site-dependent $g$-tensor introduced to explain unusual spin excitations and 
field-induced transitions in the quasi-one-dimensional Co$^{2+}$ chain. 
Our analysis has focused on the idealized octahedral limit. However, real 
materials like BaCo$_2$V$_2$O$_8$ exhibit octahedral distortions that generate three additional 
interactions \cite{churchill2024transforming}. Given the symmetry constraints imposed by the four-fold screw axis, we expect the 
qualitative features of our results to remain robust, though the quantitative 
values of the critical fields and the $g$-tensor will be modified. A systematic incorporation of these distortions represents a valuable direction for 
future study.

More broadly, the current microscopic theory  opens several future research avenues: extending the analysis to  arbitrary field directions, examining finite-temperature signatures of the two  transitions, and exploring related screw-chain systems where competing interactions may stabilize even richer quantum phases.  
Our findings establish a rare and concrete example in which long-distance Ising universality and short-distance microscopic spin textures manifest at low and high field regions, revealing how Ising-like materials can go beyond the ideal Ising model without abandoning its Ising critical behavior.

\section{Method}
We employ the density matrix renormalization group (DMRG) method \cite{White1992, schollwock2011density} for larger system sizes using the ITensor library \cite{fishman2022itensor}. The method is typically performed with a bond dimension larger than \(1000\) and a cutoff \(\varepsilon < 10^{-10} \) \cite{Sorensen2023PRRl, bhullar2025field}. We perform both the open boundary condition (OBC) and periodic boundary condition (PBC), i.e. \(\mathbf{S}_{N + 1} = \mathbf{S}_{1}\) with $N=160 - 800$. Our results are not sensitive to the boundary conditions, but the choice of the initial wavefunction 
for fields near the critical point is important. We choose several different initial wavefunctions and choose the lowest energy state throughout the entire field region. We also study the classical model to understand the impact of the bond-dependent interactions. 

\hskip 0.5cm

\section{Acknowledgments} 
This work is supported by the NSERC Discovery Grant No. 2022-04601 and NSERC CREATE program No. 575280-2023. H. Y. K. acknowledges support from the Canada Research Chairs Program No. CRC-2019-00147.
This research was enabled in part by support provided by Compute Ontario,  Calcul Québec, and the Digital Research Alliance of Canada.

\bibliography{references} 

\end{document}

%% file: macros.tex
\usepackage{txfonts}
\usepackage{graphicx}
\usepackage{dcolumn}
\usepackage{bm}
\usepackage{array}
\usepackage[dvipsnames]{xcolor}
\usepackage[%
    colorlinks=true,
    pdfborder={0 0 0},
    linkcolor=blue,
    citecolor=black
]{hyperref}  
\usepackage{chngpage}
\usepackage{appendix}

\usepackage{blindtext,subcaption,graphicx}
\usepackage{booktabs}
\usepackage{caption}
\usepackage[T1]{fontenc}
\usepackage{lmodern}

\definecolor{Dark}{gray}{0.2}
\definecolor{MedDark}{gray}{0.4}
\definecolor{Medium}{gray}{0.6}
\definecolor{Light}{gray}{0.8}
\definecolor{darkred}{rgb}{0.55, 0.0, 0.0}
\definecolor{darkslateblue}{rgb}{0.28, 0.24, 0.55}
\definecolor{royalblue(web)}{rgb}{0.25, 0.41, 0.88}
\usepackage{graphicx,float}
\usepackage{amsmath}

\usepackage{amssymb}
\hypersetup{
    citecolor=darkred,
    linkcolor=blue,   
    urlcolor=blue}

\usepackage{bbm}

\def\be{\begin{equation}}
\def\ee{\end{equation}}
\def\beq{\begin{equation}}
\def\eeq{\end{equation}}
\def\bea{\begin{eqnarray}}
\def\eea{\end{eqnarray}}
\def\nbea{\begin{eqnarray*}}
\def\neea{\nonumber\end{eqnarray*}}
\def\bmat#1{\left(\begin{array}{#1}}
\def\emat{\end{array}\right)}
\def\bcase#1{\left\{\begin{array}{#1}}
\def\ecase{\end{array}\right.}
\def\bmini#1{\begin{minipage}{#1\textwidth}}
\def\emini{\end{minipage}}

\usepackage{wrapfig}
\graphicspath{{figures/}}
\usepackage{enumerate}

%% file: references.bib
@article{Ising1925,
  author  = {Ising, Ernst},
  title   = {Beitrag zur Theorie des Ferromagnetismus},
  journal = {Zeitschrift für Physik},
  year    = {1925},
  volume  = {31},
  number  = {1},
  pages   = {253--258},
  doi     = {10.1007/BF02980577}
}

@article{Lenz1920,
  author  = {Lenz, Wilhelm},
  title   = {Beiträge zum Verständnis der magnetischen Eigenschaften in festen Körpern},
  journal = {Physikalische Zeitschrift},
  year    = {1920},
  volume  = {21},
  pages   = {613--615}
}

@article{Pfeuty1970,
  author  = {Pfeuty, Pierre},
  title   = {The one-dimensional Ising model with a transverse field},
  journal = {Annals of Physics},
  volume  = {57},
  number  = {1},
  pages   = {79--90},
  year    = {1970},
  doi     = {10.1016/0003-4916(70)90270-8}
}

@article{Jensen1976,
  author  = {Jensen, D. J. and Touborg, P. E.},
  title   = {Ferromagnetic ordering in LiHoF4},
  journal = {Physical Review B},
  volume  = {13},
  number  = {5},
  pages   = {2108--2115},
  year    = {1976},
  doi     = {10.1103/PhysRevB.13.2108}
}

@article{Kogut1979,
  author  = {Kogut, John B.},
  title   = {An introduction to lattice gauge theory and spin systems},
  journal = {Reviews of Modern Physics},
  volume  = {51},
  number  = {4},
  pages   = {659--713},
  year    = {1979},
  doi     = {10.1103/RevModPhys.51.659}
}

@book{Sachdev2011,
  author    = {Sachdev, Subir},
  title     = {Quantum Phase Transitions},
  edition   = {2},
  publisher = {Cambridge University Press},
  address   = {Cambridge},
  year      = {2011},
  isbn      = {978-0521514682}
}

@article{Fava2020,
  author  = {Fava, Massimo and Coldea, Radu and Parameswaran, S. A.},
  title   = {Glide symmetry breaking and microscopic models for the zigzag quantum Ising material CoNb2O6},
  journal = {Proceedings of the National Academy of Sciences},
  volume  = {117},
  number  = {42},
  pages   = {25219--25226},
  year    = {2020},
  doi     = {10.1073/pnas.2009389117}
}

@article{Woodland2023,
  author  = {Woodland, L. and Hong, T. and Kim, J. and others},
  title   = {Excitations of the quantum Ising chain CoNb2O6 in a low transverse field},
  journal = {Physical Review B},
  volume  = {108},
  number  = {18},
  pages   = {184417},
  year    = {2023},
  doi     = {10.1103/PhysRevB.108.184417}
}

@article{Gallegos2024,
  author  = {Gallegos, C. A. and Chernyshev, A. L.},
  title   = {Magnon interactions in the quantum paramagnetic phase of CoNb2O6},
  journal = {Physical Review B},
  volume  = {109},
  number  = {2},
  pages   = {024409},
  year    = {2024},
  doi     = {10.1103/PhysRevB.109.024409}
}

@article{Zamolodchikov1989,
  author  = {Zamolodchikov, A. B.},
  title   = {Integrals of motion and S-matrix of the (scaled) T = Tc Ising model with magnetic field},
  journal = {International Journal of Modern Physics A},
  volume  = {4},
  number  = {16},
  pages   = {4235--4248},
  year    = {1989},
  doi     = {10.1142/S0217751X8900176X}
}

@article{Kimura2007,
  author  = {Kimura, S. and Okunishi, K. and Hagiwara, M. and He, Z. and Kindo, K. and Taniyama, T. and Itoh, M.},
  title   = {Magnetic-Field-Induced Phase Transition in the Quasi-One-Dimensional Ising-Like Antiferromagnet BaCo2V2O8},
  journal = {Physical Review Letters},
  volume  = {99},
  number  = {8},
  pages   = {087602},
  year    = {2007},
  doi     = {10.1103/PhysRevLett.99.087602}
}

@article{White1992,
  author  = {White, Steven R.},
  title   = {Density matrix formulation for quantum renormalization groups},
  journal = {Physical Review Letters},
  volume  = {69},
  number  = {19},
  pages   = {2863--2866},
  year    = {1992},
  doi     = {10.1103/PhysRevLett.69.2863}
}

@article{Matsuda2025RMP,
  title = {Kitaev quantum spin liquids},
  author = {Matsuda, Yuji and Shibauchi, Takasada and Kee, Hae-Young},
  journal = {Rev. Mod. Phys.},
  volume = {97},
  issue = {4},
  pages = {045003},
  numpages = {61},
  year = {2025},
  month = {Dec},
  publisher = {American Physical Society},
  doi = {10.1103/3m4m-3v59},
  url = {https://link.aps.org/doi/10.1103/3m4m-3v59}
}

@article{Wichmann1986_BaCo2V2O8,
  author  = {R. Wichmann and H. M{\"u}ller-Buschbaum},
  title   = {Neue Verbindungen mit SrNi2V2O8-Struktur: BaCo2V2O8 und BaMg2V2O8},
  journal = {Zeitschrift f{\"u}r Anorganische und Allgemeine Chemie},
  volume  = {534},
  pages   = {153--158},
  year    = {1986},
  doi     = {10.1002/zaac.19865340320}
}

@article{He2005crystal,
title={Crystal Growth and Magnetic Properties of BaCo2V2O8},
Author = {He, Zhangzhen and Fu, Desheng and Kyômen, Tôru and Taniyama, Tomoyasu and Itoh, Mitsuru},
Journal={Chemistry of Materials},
Year ={2005},
Volume ={17},
Number = {11},
Pages ={2924-2926},
Doi={10.1021/cm050760e},
Eprint ={https://doi.org/10.1021/cm050760e},
Url ={https://doi.org/10.1021/cm050760e},
URLs ={https://doi.org/10.1021/cm050760e}
}

@article{Kawasaki2011PRB,
  title = {Magnetic structure and spin dynamics of the quasi-one-dimensional spin-chain antiferromagnet ${\mathrm{BaCo}}_{2}{\mathrm{V}}_{2}{\mathrm{O}}_{8}$},
  author = {Kawasaki, Yu and Gavilano, Jorge L. and Keller, Lukas and Schefer, J\"urg and Christensen, Niels Bech and Amato, Alex and Ohno, Takashi and Kishimoto, Yutaka and He, Zhangzhen and Ueda, Yutaka and Itoh, Mitsuru},
  journal = {Phys. Rev. B},
  volume = {83},
  issue = {6},
  pages = {064421},
  numpages = {6},
  year = {2011},
  month = {Feb},
  publisher = {American Physical Society},
  doi = {10.1103/PhysRevB.83.064421},
  url = {https://link.aps.org/doi/10.1103/PhysRevB.83.064421}
}

@article{Niesen2013PRB,
  title = {Magnetic phase diagrams, domain switching, and quantum phase transition of the quasi-one-dimensional Ising-like antiferromagnet BaCo${}_{2}$V${}_{2}$O${}_{8}$},
  author = {Niesen, S. K. and Kolland, G. and Seher, M. and Breunig, O. and Valldor, M. and Braden, M. and Grenier, B. and Lorenz, T.},
  journal = {Phys. Rev. B},
  volume = {87},
  issue = {22},
  pages = {224413},
  numpages = {9},
  year = {2013},
  month = {Jun},
  publisher = {American Physical Society},
  doi = {10.1103/PhysRevB.87.224413},
  url = {https://link.aps.org/doi/10.1103/PhysRevB.87.224413}
}

@article{He2005Field,
  author  = {Z. He and T. Taniyama and T. Ky{\^o}men and M. Itoh},
  title   = {Field-induced order-disorder transition in the quasi-one-dimensional anisotropic antiferromagnet {BaCo2V2O8}},
  journal = {Physical Review B},
  volume  = {72},
  pages   = {172403},
  year    = {2005},
  doi     = {10.1103/PhysRevB.72.172403}
}

@article{He2006Anisotropy,
  author  = {Z. He and T. Taniyama and M. Itoh},
  title   = {Large magnetic anisotropy in the quasi-one-dimensional system {BaCo2V2O8}},
  journal = {Applied Physics Letters},
  volume  = {88},
  pages   = {132504},
  year    = {2006},
  doi     = {10.1063/1.2189913}
}

@article{Ideta2013NMR,
  author  = {Y. Ideta and Y. Kawasaki and Y. Kishimoto and T. Ohno
             and Y. Michihiro and Z. He and Y. Ueda and M. Itoh},
  title   = {{$^{51}$V-NMR} study of the quasi-one-dimensional antiferromagnet {BaCo2V2O8}},
  journal = {Journal of the Korean Physical Society},
  volume  = {63},
  number  = {3},
  pages   = {739--742},
  year    = {2013},
  doi     = {10.3938/jkps.63.739}
}

@article{Mansson2012PhysProcedia,
  author  = {M. M{\aa}nsson and K. Pr{\v s}a and J. Sugiyama and T. Goko
             and C. Baines and A. Amato and F. L. Pratt and Z. He
             and M. Itoh},
  title   = {Microscopic Magnetic Nature of the Quasi-one-Dimensional
             Antiferromagnet BaCo2V2O8},
  journal = {Physica Procedia},
  volume  = {30},
  pages   = {146--150},
  year    = {2012},
  doi     = {10.1016/j.phpro.2012.04.028}
}

@inproceedings{Kawasaki2014JPSCP,
  author    = {Y. Kawasaki and Y. Ideta and Y. Kishimoto and T. Ohno
               and Y. Michihiro and Z. He and Y. Ueda and M. Itoh},
  title     = {Antiferromagnetic State in the Quasi-One-Dimensional BaCo2V2O8:
               $^{51}$V-NMR Study on a Single Crystal},
  booktitle = {Journal of the Physical Society of Japan Conference Proceedings},
  volume    = {3},
  pages     = {014001},
  year      = {2014},
  doi       = {10.7566/JPSCP.3.014001}
}

@article{Wang2018PRL_QC,
  author  = {Zhe Wang and M. Schmidt and A. Günther and N. M. Bruckner
             and M. Baenitz and Y. Skourski and J. Wosnitza
             and H. Berger and V. Zapf and B. Normand and Z. Weichselbaum
             and I. Affleck and F. Faison and P. Lejay and B. Lake},
  title   = {Quantum Criticality of an Ising-Like Spin-1/2 Antiferromagnetic Chain
             in a Transverse Magnetic Field},
  journal = {Physical Review Letters},
  volume  = {120},
  number  = {20},
  pages   = {207205},
  year    = {2018},
  doi     = {10.1103/PhysRevLett.120.207205}
}

@article{Faure2019PRL_ConfinedSpinons,
  author  = {Q. Faure and S. Takayoshi and S. Petit and B. Lake
             and S. Raymond and M. Boehm and P. Lejay and V. Simonet
             and T. Giamarchi and B. Grenier},
  title   = {From Confined Spinons to Emergent Fermions: Observing the Spinon
             Confinement–Deconfinement Crossover in the Ising-like Chain BaCo2V2O8},
  journal = {Physical Review Letters},
  volume  = {123},
  pages   = {027204},
  year    = {2019},
  doi     = {10.1103/PhysRevLett.123.027204}
}

@article{Faure2021PRR,
  author  = {Quentin Faure and Shintaro Takayoshi and Béatrice Grenier and Sylvain Petit
             and Stéphane Raymond and Martin Boehm and Pascal Lejay and Thierry Giamarchi
             and Virginie Simonet},
  title   = {Solitonic excitations in the Ising anisotropic chain {BaCo$_2$V$_2$O$_8$}
             under large transverse magnetic field},
  journal = {Physical Review Research},
  volume  = {3},
  pages   = {043227},
  year    = {2021},
  doi     = {10.1103/PhysRevResearch.3.043227},
  eprint  = {2107.02487},
  archivePrefix = {arXiv},
  primaryClass = {cond-mat.str-el}
}

@article{Okutani2021JPSJ,
  author  = {Akira Okutani and Hiroaki Onishi and Shojiro Kimura and Tetsuya Takeuchi
             and Takanori Kida and Michiyasu Mori and Atsushi Miyake
             and Masashi Tokunaga and Koichi Kindo and Masayuki Hagiwara},
  title   = {Spin Excitations of the S = 1/2 One-Dimensional Ising-Like Antiferromagnet
             BaCo2V2O8 in Transverse Magnetic Fields},
  journal = {Journal of the Physical Society of Japan},
  volume  = {90},
  number  = {4},
  pages   = {044704},
  year    = {2021},
  doi     = {10.7566/JPSJ.90.044704}
}

@article{Sorensen2021PRX,
  title = {{Heart of Entanglement: Chiral, Nematic, and Incommensurate Phases in the Kitaev-Gamma Ladder in a Field}},
  author = {S\o{}rensen, Erik S. and Catuneanu, Andrei and Gordon, Jacob S. and Kee, Hae-Young},
  journal = {Phys. Rev. X},
  volume = {11},
  issue = {1},
  pages = {011013},
  numpages = {23},
  year = {2021},
  month = {Jan},
  publisher = {American Physical Society},
  doi = {10.1103/PhysRevX.11.011013},
  url ={https://link.aps.org/doi/10.1103/PhysRevX.11.011013}
}

@article{Rouso2024RoPP,
    author = {Rousochatzakis, I and Perkins, N and Luo, Q and Kee, H.-Y.},
   journal = {Reports on Progress in Physics},
   volume = {87},
   number ={2},
    pages ={026502},
    year = {2024},
doi = {10.1088/1361-6633/ad208d},
url = {https://dx.doi.org/10.1088/1361-6633/ad208d}
}

@article{Takayoshi2018PRB,
  title = {Topological transition between competing orders in quantum spin chains},
  author = {Takayoshi, Shintaro and Furuya, Shunsuke C. and Giamarchi, Thierry},
  journal = {Phys. Rev. B},
  volume = {98},
  issue = {18},
  pages = {184429},
  numpages = {8},
  year = {2018},
  month = {Nov},
  publisher = {American Physical Society},
  doi = {10.1103/PhysRevB.98.184429},
  url = {https://link.aps.org/doi/10.1103/PhysRevB.98.184429}
}

@article{Liu2020PRL,
  title = {{Kitaev Spin Liquid in $3d$ Transition Metal Compounds}},
  author = {Liu, Huimei and Chaloupka, Ji\ifmmode \check{r}\else \v{r}\fi{}\'{\i} and Khaliullin, Giniyat},
  journal = {Phys. Rev. Lett.},
  volume = {125},
  issue = {4},
  pages = {047201},
  numpages = {6},
  year = {2020},
  month = {Jul},
  publisher = {American Physical Society},
  doi = {10.1103/PhysRevLett.125.047201},
  url = {https://link.aps.org/doi/10.1103/PhysRevLett.125.047201}
}

@article{Sano2018PRB,
  title = {{Kitaev-Heisenberg Hamiltonian for high-spin ${d}^{7}$ Mott insulators}},
  author = {Sano, Ryoya and Kato, Yasuyuki and Motome, Yukitoshi},
  journal = {Phys. Rev. B},
  volume = {97},
  issue = {1},
  pages = {014408},
  numpages = {8},
  year = {2018},
  month = {Jan},
  publisher = {American Physical Society},
  doi = {10.1103/PhysRevB.97.014408},
  url = {https://link.aps.org/doi/10.1103/PhysRevB.97.014408}
}

@article{Jackeli2009PRL,
  title = {{Mott Insulators in the Strong Spin-Orbit Coupling Limit: From Heisenberg to a Quantum Compass and Kitaev Models}},
  author = {Jackeli, G. and Khaliullin, G.},
  journal = {Phys. Rev. Lett.},
  volume = {102},
  issue = {1},
  pages = {017205},
  numpages = {4},
  year = {2009},
  month = {Jan},
  publisher = {American Physical Society},
  doi = {10.1103/PhysRevLett.102.017205},
  url = {http://link.aps.org/doi/10.1103/PhysRevLett.102.017205}
}

@article{Rau2014,
  title = {{Generic Spin Model for the Honeycomb Iridates beyond the Kitaev Limit}},
  author = {Rau, Jeffrey G. and Lee, Eric Kin-Ho and Kee, Hae-Young},
  journal = {Phys. Rev. Lett.},
  volume = {112},
  issue = {7},
  pages = {077204},
  numpages = {5},
  year = {2014},
  month = {Feb},
  publisher = {American Physical Society},
  doi = {10.1103/PhysRevLett.112.077204},
  url = {http://link.aps.org/doi/10.1103/PhysRevLett.112.077204}
}

@article{morris2021np,
	title = {{Duality and domain wall dynamics in a twisted Kitaev chain}},
	volume = {17},
	copyright = {2021 The Author(s), under exclusive licence to Springer Nature Limited},
	issn = {1745-2481},
	url = {https://www.nature.com/articles/s41567-021-01208-0},
	doi = {10.1038/s41567-021-01208-0},
	number = {7},
	urldate = {2023-04-11},
	journal = {Nature Physics},
	author = {Morris, C. M. and Desai, Nisheeta and Viirok, J. and Hüvonen, D. and Nagel, U. and Rõõm, T. and Krizan, J. W. and Cava, R. J. and McQueen, T. M. and Koohpayeh, S. M. and Kaul, Ribhu K. and Armitage, N. P.},
	month = jul,
	year = {2021},
	pages = {832--836},
}

@article{Sorensen2023PRRl,
  title = {{Field-induced chiral soliton phase in the Kitaev spin chain}},
  author = {S\o{}rensen, Erik S. and Gordon, Jacob and Riddell, Jonathon and Wang, Tianyi and Kee, Hae-Young},
  journal = {Phys. Rev. Res.},
  volume = {5},
  issue = {1},
  pages = {L012027},
  numpages = {7},
  year = {2023},
  month = {Feb},
  publisher = {American Physical Society},
  doi = {10.1103/PhysRevResearch.5.L012027},
  url = {https://link.aps.org/doi/10.1103/PhysRevResearch.5.L012027}
}

@article{fishman2022itensor,
  title={The ITensor software library for tensor network calculations},
  author={Fishman, Matthew and White, Steven and Stoudenmire, Edwin Miles},
  journal={SciPost Physics Codebases},
  pages={004},
  year={2022},
  url={https://doi.org/10.48550/arXiv.2007.14822}
}

@article{schollwock2011density,
  title={The density-matrix renormalization group in the age of matrix product states},
  author={Schollw{\"o}ck, Ulrich},
  journal={Annals of physics},
  volume={326},
  number={1},
  pages={96--192},
  year={2011},
  publisher={Elsevier},
url={https://doi.org/10.1016/j.aop.2010.09.012}
}

@article{churchill2024transforming,
  title={Transforming from Kitaev to Disguised Ising Chain: Application to CoNb 2 O 6},
  author={Churchill, Derek and Kee, Hae-Young},
  journal={Physical Review Letters},
  volume={133},
  number={5},
  pages={056703},
  year={2024},
  publisher={APS},
url={https://doi.org/10.1103/PhysRevLett.133.056703}
}

@article{coldea2010quantum,
  title={Quantum criticality in an Ising chain: experimental evidence for emergent E8 symmetry},
  author={Coldea, Radu and Tennant, DA and Wheeler, EM and Wawrzynska, E and Prabhakaran, D and Telling, M and Habicht, K and Smeibidl, P and Kiefer, K},
  journal={Science},
  volume={327},
  number={5962},
  pages={177--180},
  year={2010},
  publisher={American Association for the Advancement of Science},
url={
https://doi.org/10.48550/arXiv.1103.3694}
}

@article{zou20218,
  title={E 8 spectra of quasi-one-dimensional antiferromagnet BaCo 2 V 2 O 8 under transverse field},
  author={Zou, Haiyuan and Cui, Yi and Wang, Xiao and Zhang, Z and Yang, J and Xu, Guangyong and Okutani, A and Hagiwara, M and Matsuda, M and Wang, G and others},
  journal={Physical review letters},
  volume={127},
  number={7},
  pages={077201},
  year={2021},
  publisher={APS},
url={https://doi.org/10.1103/PhysRevLett.127.077201}
}

@article{halati2023repulsively,
  title={Repulsively bound magnon excitations of a spin-1 2 XXZ chain in a staggered transverse field},
  author={Halati, Catalin-Mihai and Wang, Zhe and Lorenz, Thomas and Kollath, Corinna and Bernier, Jean-S{\'e}bastien},
  journal={Physical Review B},
  volume={108},
  number={22},
  pages={224429},
  year={2023},
  publisher={APS},
url={
https://doi.org/10.48550/arXiv.2306.14742
}
}

@article{bhullar2025field,
  title={Field-induced ordered phases in anisotropic spin-1 2 Kitaev chains},
  author={Bhullar, Mandev and Xu, Haoting and Kee, Hae-Young},
  journal={Physical Review B},
  volume={111},
  number={10},
  pages={104439},
  year={2025},
  publisher={APS},
url = {https://doi.org/10.1103/PhysRevB.111.104439}
}

@article{faure2018topological,
  title={Topological quantum phase transition in the Ising-like antiferromagnetic spin chain BaCo2V2O8},
  author={Faure, Quentin and Takayoshi, Shintaro and Petit, Sylvain and Simonet, Virginie and Raymond, St{\'e}phane and Regnault, Louis-Pierre and Boehm, Martin and White, Jonathan S and M{\aa}nsson, Martin and R{\"u}egg, Christian and others},
  journal={Nature Physics},
  volume={14},
  number={7},
  pages={716--722},
  year={2018},
  publisher={Nature Publishing Group UK London},
url={https://doi.org/10.1038/s41567-018-0126-8}
}

@article{kimura2013collapse,
  title={Collapse of magnetic order of the quasi one-dimensional Ising-like antiferromagnet BaCo2V2O8 in transverse fields},
  author={Kimura, Shojiro and Okunishi, Koichi and Hagiwara, Masayuki and Kindo, Koichi and He, Zhangzhen and Taniyama, Tomoyasu and Itoh, Mitsuru and Koyama, Keiichi and Watanabe, Kazuo},
  journal={Journal of the Physical Society of Japan},
  volume={82},
  number={3},
  pages={033706},
  year={2013},
  publisher={The Physical Society of Japan},
url={https://doi.org/10.7566/JPSJ.82.033706}
}

@article{giamarchi1988theory,
  title={Theory of spin-anisotropic electron-electron interactions in quasi-one-dimensional metals},
  author={Giamarchi, T and Schulz, HJ},
  journal={Journal de Physique},
  volume={49},
  number={5},
  pages={819--835},
  year={1988},
  publisher={Soci{\'e}t{\'e} fran{\c{c}}aise de physique},
url={	https://doi.org/10.1051/jphys:01988004905081900
}
}

@article{Lecheminant:2002va,
    author = "Lecheminant, P. and Gogolin, Alexander O. and Nersesyan, Alesander A.",
    title = "{Criticality in selfdual sine-Gordon models}",
    eprint = "cond-mat/0203294",
    archivePrefix = "arXiv",
    doi = "10.1016/S0550-3213(02)00474-1",
    journal = "Nucl. Phys. B",
    volume = "639",
    pages = "502--523",
    year = "2002",
url={https://doi.org/10.1016/S0550-3213(02)00474-1}
}

@article{fava2020glide,
  title={Glide symmetry breaking and Ising criticality in the quasi-1D magnet CoNb2O6},
  author={Fava, Michele and Coldea, Radu and Parameswaran, SA},
  journal={Proceedings of the National Academy of Sciences},
  volume={117},
  number={41},
  pages={25219--25224},
  year={2020},
  publisher={National Academy of Sciences},
url = {
https://doi.org/10.1073/pnas.2007986117}
}

@article{Liu2018PRB,
  title = {Pseudospin exchange interactions in ${d}^{7}$ cobalt compounds: Possible realization of the Kitaev model},
  author = {Liu, Huimei and Khaliullin, Giniyat},
  journal = {Phys. Rev. B},
  volume = {97},
  issue = {1},
  pages = {014407},
  numpages = {12},
  year = {2018},
  month = {Jan},
  publisher = {American Physical Society},
  doi = {10.1103/PhysRevB.97.014407},
  url = {https://link.aps.org/doi/10.1103/PhysRevB.97.014407}
}

@article{Winter_2022,
doi = {10.1088/2515-7639/ac94f8},
url = {https://dx.doi.org/10.1088/2515-7639/ac94f8},
year = {2022},
month = {oct},
publisher = {IOP Publishing},
volume = {5},
number = {4},
pages = {045003},
author = {Winter, Stephen M},
title = {Magnetic couplings in edge-sharing high-spin d7 compounds},
journal = {Journal of Physics: Materials}
}

@article{cui2019quantum,
  title={Quantum criticality of the Ising-like screw chain antiferromagnet SrCo2V2O8 in a transverse magnetic field},
  author={Cui, Y and Zou, H and Xi, N and He, Zhangzhen and Yang, YX and Shu, L and Zhang, GH and Hu, Z and Chen, T and Yu, Rong and others},
  journal = {Phys. Rev. Lett.},
  volume = {123},
  issue = {6},
  pages = {067203},
  numpages = {6},
  year = {2019},
  month = {Aug},
  publisher = {American Physical Society},
  doi = {10.1103/PhysRevLett.123.067203},
  url = {https://link.aps.org/doi/10.1103/PhysRevLett.123.067203}
}

@article{mitra2025quantum,
  title = {Quantum criticality and universality in the stationary state of the long-range Kitaev model},
  author = {Mitra, Akash and Paul, Sanku and Srivastava, Shashi C. L.},
  journal = {Phys. Rev. B},
  volume = {111},
  issue = {10},
  pages = {104308},
  numpages = {11},
  year = {2025},
  month = {Mar},
  publisher = {American Physical Society},
  doi = {10.1103/PhysRevB.111.104308},
  url = {https://link.aps.org/doi/10.1103/PhysRevB.111.104308}
}

@article{Wu1976_Ising_SpinSpin,
  author  = {Tai Tsun Wu and Barry M. McCoy and Craig A. Tracy and Eytan Barouch},
  title   = {Spin-spin correlation functions for the two-dimensional Ising model:
             Exact theory in the scaling region},
  journal = {Physical Review B},
  volume  = {13},
  number  = {1},
  pages   = {316--374},
  year    = {1976},
  doi     = {10.1103/PhysRevB.13.316}
}

@article{Montroll1963_Ising_Correlations,
  author  = {Elliott W. Montroll and Renfrey B. Potts and John C. Ward},
  title   = {Correlations and Spontaneous Magnetization of the Two-Dimensional Ising Model},
  journal = {Journal of Mathematical Physics},
  volume  = {4},
  pages   = {308--322},
  year    = {1963},
  doi     = {10.1063/1.1703955}
}
